\documentclass[seceq]{ptptex}

\title{
Supersymmetric Construction of three-dimensional isospectral systems}

\author{
Yoshihide \textsc{Yamada}%
}

\inst{
Graduate School of Human and Environment Studies,\\
 Kyoto University, Kyoto 606-8501, Japan}

\abst{
Three-dimensional isospectral systems are constructed using the framework of supersymmetric quantum mechanics.
In case the supercharge of first order in momentum is used, it is proved that the constructed systems reduce to a trivial generalization of one-dimensional systems.
In case the second order supercharge is used, nontrivial three-dimensional isospectral systems are constructed.
}

\begin{document}

\maketitle

\section{Introduction}

The concept of supersymmetry arose in the study of elementary particle physics.
The term ``supersymmetry'' was originally used in reference to the symmetry of bosons and fermions, which behave according to different statistical laws.
Supersymmetry is widely regarded as a necessary concept for unifying all the elementary forces.
The concept of supersymmetry was originally formulated in the framework of field theory.
However, it is not easy to analyze the results from supersymmetry (that is, whether or not a model exhibits the breaking of supersymmetry) in field theory.
For this reason, supersymmetric quantum mechanics was proposed \cite{-a}.
Although the original purpose of formulating supersymmetric quantum mechanics was to help obtain an understanding of supersymmetric quantum field theory, once this formulation of quantum mechanics was made, many interesting applications were found.
Supersymmetric quantum mechanics is actually a reformulation of Darboux-Crum transformations and the factorization method \cite{-b}.
Darboux-Crum transformations are techniques of generating a new Hamiltonian and its eigenfunctions from an existing Hamiltonian and its known eigenfunctions.
The new spectrum is almost identical to the original \cite{1a,1b}.
The factorization method is a technique of solving Schr\"odinger equation by factorizing the Hamiltonian, which is the second order differential operator with respect to the spatial variables in position space, into the first order operators \cite{1c}.

In quantum mechanics, supersymmetry refers to a relationship of isospectrality of two systems.
More precisely, supersymmetric quantum mechanics is a formulation of quantum mechanics involving pairs of Hamiltonians (so called ``partner Hamiltonians'') whose eigenvalues are identical (except for a possible ground state).
The existence of such (non-trivial) pairs is possible because there is not a one-to-one correspondence between a spectrum and a potential.
Indeed, for a given spectra there will be many potentials from which it can be derived.
Supersymmetric quantum mechanics provides a method to construct pairs of isospectral systems \cite{00a}.

An operator called intertwiner plays a central role in constructing isospectral systems.
Intertwiner was first introduced as factors of the Hamiltoinans, and therefore they were of first order in the momentum.
Later its generalization to the higher order in the momentum was proposed, which made it possible to construct other classes of isospectral systems \cite{0a,0b,0c}.

However, most investigations of systems on supersymmetric quantum mechanics and its higher order generalization are restricted to one-dimensional systems (or spherically symmetric three-dimensional systems).
Two-dimensional generalization of supersymmetric quantum mechanics using first order intertwiner was proposed, which established an isospectrality of matrix Hamiltonian and a pair of scalar Hamiltonians \cite{00ab}.
Two-dimensional isospectral systems without matrix Hamiltonian were constructed using the second order intertwiner \cite{00b,00c}.
In this paper, we construct three-dimensional isospectral systems using supersymmetric quantum mechanics and its second order generalization.
Two-dimensional isospectral systems based on first order intertwiner is also obtained from our result in three-dimensional case by taking the $z$-component of relevant quantity to zero.

In section 2, we review the one-dimensional supersymmetric quantum mechanics and its second order generalization emphasizing the scheme for constructing isospectral systems.
In section 3, we investigate three-dimensional generalization with first order intertwiner.
In section4, we investigate three-dimensional generalization with second order intertwiner.

\section{Review of one-dimensional case}

\subsection{Supersymmetric quantum mechanics}

Let us see how one-dimensional isospectral systems are constructed using the framework of supersymmetric quantum mechanics.
We consider a pair of nonrelativistic Hamiltonians;
\begin{align}
H & = -\partial^2 +V(x), \\
\widetilde{H} &= -\partial^2 + \widetilde{V}(x),
\end{align}
where $\partial \equiv d/dx$.
If these Hamiltonians satisfy the following relations;
\begin{align}
AH &= \widetilde{H}A, \label{1}\\ 
A^\dag\widetilde{H} &= HA^\dag,
\end{align}
which are called ``intertwining relations,'' then these systems are isospectral (except for zero mode of $A$ or $A^\dag$).
This is because by multiplying $A$ and $A^\dag$ to
\begin{align}
H\psi_E &= E\psi_E, \\
\widetilde{H}\widetilde{\psi}_{\widetilde{E}} &= \widetilde{E}\widetilde{\psi}_{\widetilde{E}}
\end{align}
respectively, we obtain
\begin{align}
\widetilde{H}(A\psi_E) &= E(A\psi_E), \\
H(A^\dag\widetilde{\psi}_{\widetilde{E}}) &= \widetilde{E}(A^\dag\widetilde{\psi}_{\widetilde{E}}).
\end{align}
From these relations, we see that $H$ and $\widetilde{H}$ are basically isospectral (except that it is possible for the ground state of $H$ or $\widetilde{H}$ to be missing), and their eigenfunctions are related as
\begin{align}
\widetilde{\psi}_E &\propto A\psi_E, \\
\psi_E &\propto A^\dag\widetilde{\psi}_E.
\end{align}

In the case the intertwiner of the form
\begin{align}
A = v(x)\partial + w(x)
\end{align}
is substituted to Eq.\eqref{1}, we obtain the following equations for $v, w, V$ and $\widetilde{V}$;
\begin{align}
&v^\prime=0, \label{02}\\
&v(\widetilde{V}-V) - v^{\prime\prime} -2w^\prime =0, \label{2} \\
&w(\widetilde{V}-V) - vV^\prime - w^{\prime\prime} =0, \label{3}
\end{align}
where prime denotes derivative with respect to the variable of the function.
From Eq.\eqref{02} we can set $v=1$ without losing generality because $A$ is determined only up to a multiplicative constant, as is evident in Eq.\eqref{1}.
Then from Eq.\eqref{2} we obtain
\begin{align}
\widetilde{V} - V = 2w^\prime,
\end{align}
and substituting this to Eq.\eqref{3}, we find
\begin{align}
V &= w^2 - w^\prime +c, \\
\widetilde{V} &= w^2 + w^\prime +c,
\end{align}
where $c$ is an integration constant.
The function $w(x)$ remains arbitrary.
By choosing the function $w(x)$, isospectral systems $V, \widetilde{V}$ are obtained.
The function $w(x)$ is called the superpotential in supersymmetric quantum mechanics (though it corresponds to the derivative of the superpotential in supersymmetric quantum field theory).
Products of $A$ and $A^\dag$ are related to the Hamiltonians as follows,
\begin{align}
A^\dag A &= H - c, \label{factorization1}\\ 
A A^\dag &= \widetilde{H} - c. \label{factorization2}
\end{align}
As noted above, the ground state of $H$ or $\widetilde{H}$ could be missing from the spectrum of the other.
This is seen as follows:
By defining the zero-mode functions of intertwiner $A$ and $A^\dag$ as
\begin{align}
A\phi^- \equiv 0, \\
A^\dag\phi^+ \equiv 0,
\end{align}
we have
\begin{align}
H\phi^- = (A^\dag A + c)\phi^- = c\phi^- , \\
\widetilde{H}\phi^+ = (AA^\dag  + c)\phi^+ = c\phi^+ .
\end{align}
If $\phi^\pm$ are regular and normalizable, they are eigenfunctions of $H$ or $\widetilde{H}$ belonging to eigenvalue $c$ which are ground states because from Eqs.\eqref{factorization1} and \eqref{factorization2} $H-c$ and $\widetilde{H}-c$ are positive semi-definite($E-c\ge0$).
Yet as $\phi^\pm$ are written in terms of $w(x)$ as
\begin{align}
\phi^\pm = \exp\left( \pm\int\!\!dx\,w(x) \right),
\end{align}
they are not normalizable simultaneously.
Thus if $\phi^-$ ($\phi^+$) is normalizable, the ground state of $\widetilde{H}$ ($H$) is missing.
And if neither of $\phi^\pm$ are normalizable, spectra of $H$ and $\widetilde{H}$ are identical and the missing of ground state does not occur (so called broken SUSY).
The normalizability of $\phi^\pm$ is decided from aymptotic behavior of superpotential $w(x)$\cite{-a}.

By defining superhamiltonian $\boldsymbol{H}$, supercharges $\boldsymbol{Q}$ and $\boldsymbol{Q}^\dag$ as
\begin{align}
\boldsymbol{H} 
&\equiv 
\begin{pmatrix}
H-c & 0 \\
0 & \widetilde{H}-c
\end{pmatrix}, \\
\boldsymbol{Q} 
&\equiv 
\begin{pmatrix}
0 & 0 \\
A & 0
\end{pmatrix},
\quad
\boldsymbol{Q}^\dag 
\equiv 
\begin{pmatrix}
0 & A^\dag \\
0 & 0
\end{pmatrix},
\end{align}
we find the following supersymmetry algebra;
\begin{align}
&[\boldsymbol{H},\boldsymbol{Q}] = [\boldsymbol{H},\boldsymbol{Q}^\dag] = \boldsymbol{0}, \label{susy_algebra1}\\
&\{ \boldsymbol{Q},\boldsymbol{Q}^\dag \} = \boldsymbol{H}, \label{anti} \\
&\boldsymbol{Q}^2 =\boldsymbol{Q}^{\dag2} = \boldsymbol{0},\label{susy_algebra3}
\end{align}
where $\boldsymbol{0}$ is the $2\times2$ null matrix.
This is why this formulation is called supersymmetric quantum mechanics.

\subsection{Second order generalization}

Let us now consider a second order intertwiner \cite{d,b};
\begin{align}
A = u(x)\partial^2 + v(x)\partial + w(x).
\end{align}
Substituting this to Eq.\eqref{1}, we obtain the following equations for $u, v, w, V$ and $\widetilde{V}$;
\begin{align}
&u^\prime = 0,\label{04} \\
&u(\widetilde{V}-V) -u^{\prime\prime} - 2v^\prime =0, \label{4} \\
&v(\widetilde{V}-V) - 2uV^\prime -v^{\prime\prime} - 2w^\prime =0, \label{5}\\
&w(\widetilde{V}-V) - uV^{\prime\prime} - vV^\prime - w^{\prime\prime} =0.\label{6}
\end{align}
From Eq.\eqref{04} we can set $u=1$ without losing generality for the same reason as the previous case.
Then from Eq.\eqref{4} we obtain
\begin{align}
\widetilde{V} - V = 2v^\prime,
\end{align}
and substituting this to Eqs.\eqref{5} and \eqref{6} we have
\begin{align}
V + w &= \frac{1}{2}(v^2 - v^\prime) + c, \label{7}\\
(V + w)^{\prime\prime} &= 2(V + w)v^\prime -2v^\prime V -vV^\prime,
\end{align}
where $c$ is an integration constant.
Eliminating $(V+w)$ and multiplying $v$, we have
\begin{align}
(v^2V)^\prime = \left(\frac{v^4}{4} + cv^2 - v^2v^\prime + \frac{vv^{\prime\prime}}{2} - \frac{v^{\prime2}}{4}\right)^\prime.
\end{align}
Thus we obtain
\begin{align}
V &= \frac{v^2}{4} + c - v^\prime + \frac{v^{\prime\prime}}{2v} - \frac{v^{\prime2}}{4v^2} + \frac{d}{v^2}, \\
\widetilde{V} &= \frac{v^2}{4} + c + v^\prime + \frac{v^{\prime\prime}}{2v} - \frac{v^{\prime2}}{4v^2} + \frac{d}{v^2},
\end{align}
and from Eq.\eqref{7}
\begin{align}
w = \frac{v^2}{4} + \frac{v^\prime}{2} - \frac{v^{\prime\prime}}{2v} + \frac{v^{\prime2}}{4v^2} - \frac{d}{v^2},
\end{align}
where $d$ is an integration constant.
The function $v(x)$ remains arbitrary.
By choosing the function $v(x)$ isospectral systems $V, \widetilde{V}$ are obtained.
Products of $A$ and $A^\dag$ are related to the Hamiltonians as follows,
\begin{align}
A^\dag A &= (H - c)^2-d, \label{factorization1-2nd}\\
A A^\dag &= (\widetilde{H} - c)^2-d.\label{factorization2-2nd}
\end{align}
As there are two independent zero-mode functions of $A$, the spectra of $H$ and $\widetilde{H}$ could be different at most two energy levels depending on their regularity and normalizability.
This possible difference in spectra is rather delicate problem and it is usually analyzed based on Eqs.\eqref{factorization1-2nd}, \eqref{factorization2-2nd} and in terms of Darboux-Crum transformations \cite{0b,com1,com2}.

The anticommutator of $\boldsymbol{Q}$ and $\boldsymbol{Q}^\dag$ becomes
\begin{align}
\{ \boldsymbol{Q},\boldsymbol{Q}^\dag \} = \boldsymbol{H}^2 -d \boldsymbol{1},
\end{align}
where $\boldsymbol{1}$ is the $2\times2$ identity matrix.
This together with Eqs.\eqref{susy_algebra1} and \eqref{susy_algebra3} form extended nonlinear superalgebra.

\section{Three-dimensional generalization with first order intertwiner}

Here we consider the three-dimensional generalization using the first order intertwiner.
Hamiltonians are
\begin{align}
H &= - \triangle + V(\boldsymbol{r}),\label{Hamilton} \\
\widetilde{H} &= - \triangle + \widetilde{V}(\boldsymbol{r}), \label{Hamilton2}
\end{align}
where $\boldsymbol{r} = (x,y,z)$, $\triangle = \partial_x^2 + \partial_y^2 + \partial_z^2$, and the intertwiner is the form
\begin{align}
A = \boldsymbol{v}(\boldsymbol{r})\cdot\nabla + w(\boldsymbol{r}),
\end{align}
where $\boldsymbol{v}(\boldsymbol{r})=(v_x(\boldsymbol{r}),v_y(\boldsymbol{r}),v_z(\boldsymbol{r}))$.
Substituting these to 
\begin{align}
AH &= \widetilde{H}A, \label{re1}
\end{align}
we obtain the following equations for $\boldsymbol{v}, w, V$ and $\widetilde{V}$;
\begin{align}
&\partial_iv_j + \partial_jv_i = 0, \label{8}\\
&\triangle \boldsymbol{v} + 2\nabla w = (\widetilde{V}-V)\boldsymbol{v},\label{9} \\
&\boldsymbol{v}\cdot\nabla V + \triangle w = (\widetilde{V}-V)w.\label{10}
\end{align}
The general solution of Eq.\eqref{8} is
\begin{align}
\boldsymbol{v} = \boldsymbol{a}\times\boldsymbol{r} + \boldsymbol{b}, \label{10.1}
\end{align}
where $\boldsymbol{a}, \boldsymbol{b}$ are arbitrary constant vectors.
Thus the intertwiner has the form
\begin{align}
A = \boldsymbol{a}\cdot(\boldsymbol{r}\times\nabla) + \boldsymbol{b}\cdot\nabla + w.
\end{align}
Substituting Eq.\eqref{10.1} into Eqs.\eqref{9} and \eqref{10}, these become
\begin{align}
&2\nabla w = (\widetilde{V}-V)(\boldsymbol{a}\times\boldsymbol{r} + \boldsymbol{b}), \label{13}\\
&\boldsymbol{a}\cdot(\boldsymbol{r}\times\nabla V) + \boldsymbol{b}\cdot\nabla V + \triangle w = (\widetilde{V}-V)w.\label{14}
\end{align}
We find it convenient to deal with two cases, (i) $\boldsymbol{a}\neq\boldsymbol{0}$ and (ii) $\boldsymbol{a}=\boldsymbol{0}$, separately.

\ 

(i)
When $\boldsymbol{a}\neq\boldsymbol{0}$, we can take $\boldsymbol{a}$ to a unit vector without losing generality because $A$ is determined only up to a multiplicative constant.
By making rotation of coordinates, we can set $\boldsymbol{a}=\boldsymbol{e}_z$ (the unit vector in the direction of the $z$ axis);
\begin{align}
\boldsymbol{v} = \boldsymbol{e}_z\times(\boldsymbol{r}-\boldsymbol{e}_z\times\boldsymbol{b}) + b_z\boldsymbol{e}_z.
\end{align}
By translation of the origin $\boldsymbol{r} - \boldsymbol{e}_z\times\boldsymbol{b} \to \boldsymbol{r}$, this reduces to
\begin{align}
\boldsymbol{v} = \boldsymbol{e}_z\times\boldsymbol{r} + b_z\boldsymbol{e}_z,
\end{align}
which corresponds to $\boldsymbol{a}=\boldsymbol{e}_z$ and $\boldsymbol{b}=b_z\boldsymbol{e}_z$.
Thus the intertwiner becomes
\begin{align}
A= \boldsymbol{e}_z\cdot(\boldsymbol{r}\times\nabla) + b_z\partial_z + w(\boldsymbol{r}).
\end{align}
Eqs.\eqref{13} and \eqref{14} become
\begin{align}
&2\nabla w = (\widetilde{V}-V)(\boldsymbol{e}_z\times\boldsymbol{r}+b_z\boldsymbol{e}_z), \\
&\boldsymbol{e}_z\cdot(\boldsymbol{r}\times\nabla V) + b_z\partial_z V + \triangle w =(\widetilde{V}-V)w,
\end{align}
which reduces to the following equations using cylindrical coordinates $(\rho,\varphi,z)$,
\begin{align}
&2\partial_\rho w = 0, \label{010}\\
&2\partial_\varphi w = (\widetilde{V}-V)\rho^2, \label{020}\\
&2\partial_z w = (\widetilde{V}-V)b_z, \label{030}\\
&(\partial_\varphi + b_z\partial_z)V + \left(\partial_\rho^2 + \frac{1}{\rho}\partial_\rho + \frac{1}{\rho^2}\partial_\varphi^2 + \partial_z^2\right)w = (\widetilde{V}-V)w.\label{040}
\end{align}

When $b_z\neq0$, by differentiating Eqs.\eqref{020} and \eqref{030} with respect to $\rho$, we obtain a singular (on the $z$ axix) potential
\begin{align}
\widetilde{V} = V = V(\rho,\xi), \quad \xi \equiv b_z\varphi - z.
\end{align}
For single-valuedness, $V(\rho,\xi)$ must be a periodic function of $\xi$ with a period $2\pi b_z$.
Since $\widetilde{H} =H$, the intertwiner becomes a symmetry operator of this system;
\begin{align}
[H,A]=0,\quad A=\partial_\varphi + b_z\partial_z.
\end{align}
This is simply because $[A,\xi]=0$.

When $b_z=0$, Eq.\eqref{030} reduces to $\partial_z w=0$, which together with Eq.\eqref{010}, yields
\begin{align}
w=w(\varphi). \label{040a}
\end{align}
Eq.\eqref{020} can be written as follows;
\begin{align}
\widetilde{V}-V = \frac{2w^\prime(\varphi)}{\rho^2}. \label{040b}
\end{align}
By substituting Eqs.\eqref{040a} and \eqref{040b} to Eq.\eqref{040}, we have
\begin{align}
\partial_\varphi V + \frac{w^{\prime\prime}(\varphi)}{\rho^2} = \frac{2ww^\prime(\varphi)}{\rho^2}.
\end{align}
Thus we obtain
\begin{align}
V &= \frac{w^2(\varphi)-w^\prime(\varphi)}{\rho^2} + V_{\rho z}(\rho,z), \label{rho1} \\
\widetilde{V} &= \frac{w^2(\varphi)+w^\prime(\varphi)}{\rho^2} + V_{\rho z}(\rho,z) ,\label{rho2}
\end{align}
where $V_{\rho z}$ is an arbitrary function of $\rho,z$, and
\begin{align}
A= \partial_\varphi + w(\varphi). \label{050a}
\end{align}
And for single-valuedness, $w(\varphi)$ must be a periodic function with a period $2\pi$.
These potentials are singular on the $z$ axis.
To avoid the singularity, we have two means.
One is to take $w(\varphi) = \mathrm{const}$, which results in a trivial case $V=\widetilde{V}=V(\rho,z)$ and $A = \partial_\varphi + \mathrm{const}$.
The other is to consider that there is a cylindrical matter on the $z$-axis with the finite radius $\rho_0$, in such a case we may restrict the region of definition as $\rho \ge \rho_0$ and assume a suitable boundary condition on wave function at $\rho=\rho_0$.
In any way, Eqs.\eqref{rho1}, \eqref{rho2} and \eqref{050a} are trivial generalization of the one-dimensional case, because of the following reason:
Firstly the following relation is satisfied by $A$ in Eq.\eqref{050a};
\begin{align}
AH_{\rho z} = H_{\rho z}A, \label{050c}
\end{align}
where $H_{\rho z} = - \rho^{-1}\partial_\rho\rho\partial_\rho - \partial_z^2 + V_{\rho z}(\rho,z)$.
Secondly,
\begin{align}
AH_\varphi=\widetilde{H}_\varphi A,\label{050d}
\end{align}
where $H_\varphi = - \partial_\varphi^2 + w^2(\varphi) - w^\prime(\varphi)$ and $\widetilde{H}_\varphi = - \partial_\varphi^2 + w^2(\varphi) + w^\prime(\varphi)$.
By adding Eqs.\eqref{050c} and \eqref{050d} devided by $\rho^2$, we obtain
\begin{align}
A\left(\frac{H_\varphi}{\rho^2}+H_{\rho z}\right)
=\left(\frac{\widetilde{H}_\varphi}{\rho^2}+H_{\rho z}\right)A.
\end{align}
The potentials in $H_\varphi/\rho^2 + H_{\rho z}$ and $\widetilde{H}_\varphi/\rho^2 + H_{\rho z}$ are of the form Eqs.\eqref{rho1} and \eqref{rho2}

\ 

(ii)
When $\boldsymbol{a}=\boldsymbol{0}$, we can take $\boldsymbol{b}$ to a unit vector without losing generality.
By making rotation of coordinates, we can set $\boldsymbol{b}=\boldsymbol{e}_x$ (the unit vector in the direction of the $x$ axis).
Then the intertwiner becomes
\begin{align}
A = \partial_x + w(\boldsymbol{r}),
\end{align}
and Eqs.\eqref{13} and \eqref{14} become
\begin{align}
&2\nabla w = (\widetilde{V}-V)\boldsymbol{e}_x, \label{11}\\
&\partial_xV + \triangle w = (\widetilde{V}-V)w.\label{12}
\end{align}
Since $y,z$ components of Eq.\eqref{11} reduce to $\partial_yw=\partial_zw=0$, we have
\begin{align}
w=w(x). \label{wx}
\end{align}
Then $x$ component of Eq.\eqref{11} becomes
\begin{align}
\widetilde{V}-V = 2w^\prime(x).\label{Vx}
\end{align}
By substituting Eqs.\eqref{wx} and \eqref{Vx} to Eq.\eqref{12}, we have
\begin{align}
\partial_xV + w^{\prime\prime}(x) &= 2w(x)w^\prime(x).
\end{align}
Thus we obtain
\begin{align}
V &= w^2(x) - w^\prime(x) + V_{yz}(y,z), \\
\widetilde{V} &= w^2(x) + w^\prime(x) + V_{yz}(y,z),
\end{align}
where $V_{yz}(y,z)$ is an arbitrary function and
\begin{align}
A = \partial_x + w(x). \label{Ax}
\end{align}
This is a trivial generalization of one-dimensional case, because of the following reason:
Firstly the following relation is satisfied by $A$ in Eq.\eqref{Ax},
\begin{align}
AH_{yz} = H_{yz}A, \label{AHyz}
\end{align}
where $H_{yz} = - \partial_y^2 - \partial_z^2 + V_{yz}(y,z)$.
Secondly,
\begin{align}
AH_x=\widetilde{H}_xA, \label{AHx}
\end{align}
where $H_x = - \partial_x^2 + w^2(x) - w^\prime(x)$ and $\widetilde{H}_x = - \partial_x^2 + w^2(x) + w^\prime(x)$.
By adding Eqs.\eqref{AHyz} and \eqref{AHx}, we obtain
\begin{align}
A(H_x+H_{yz})=(\widetilde{H}_x+H_{yz})A.
\end{align}

\

By the results of (i) and (ii), we have proved that the three-dimensional generalization with the first order intertwiner reduces to a trivial generalization of one-dimensional case.

\section{Three-dimensional generalization with second order intertwiner}

Here we consider the three-dimensional generalization using a second order intertwiner.
Hamiltonians are Eqs.\eqref{Hamilton}, \eqref{Hamilton2}, and the intertwiner is the form
\begin{align}
A = g_{ij}(\boldsymbol{r})\partial_i\partial_j + \boldsymbol{v}(\boldsymbol{r})\cdot\nabla + w(\boldsymbol{r}). \label{16}
\end{align}
Substituting these to Eq.\eqref{re1} we obtain the following equations for $g_{ij}, \boldsymbol{v}, w, V$ and $\widetilde{V}$;
\begin{align}
&\partial_ig_{jk} + \partial_jg_{ki} + \partial_kg_{ij} = 0,\label{15} \\
&\triangle g_{ij} + \partial_iv_j + \partial_jv_i = (\widetilde{V}-V)g_{ij},\label{17} \\
&2g_{ij}\partial_jV + \triangle v_i + 2\partial_iw = (\widetilde{V}-V)v_i,\label{18} \\
&g_{ij}\partial_i\partial_jV + \boldsymbol{v}\cdot\nabla V + \triangle w = (\widetilde{V}-V)w.\label{19}
\end{align}
The general solution of Eq.\eqref{15} is
\begin{align}
g_{11} =& a_{33}x_2^2 + a_{22}x_3^2 - 2a_{23}x_2x_3 - b_{31}x_2 + b_{21}x_3 + c_{11}, \\
g_{22} =& a_{11}x_3^2 + a_{33}x_1^2 - 2a_{31}x_3x_1 - b_{12}x_3 + b_{32}x_1 + c_{22}, \\
g_{33} =& a_{22}x_1^2 + a_{11}x_2^2 - 2a_{12}x_1x_2 - b_{23}x_1 + b_{13}x_2 + c_{33}, \\
g_{12} =& - a_{33}x_1x_2 + a_{23}x_3x_1 + a_{31}x_2x_3 - a_{12}x_3^2 \nonumber\\
 &+ \frac{b_{31}}{2}x_1 - \frac{b_{32}}{2}x_2 - \frac{b_{11}-b_{22}}{2}x_3 + c_{12}, \\
g_{23} =& - a_{11}x_2x_3 + a_{31}x_1x_2 + a_{12}x_3x_1 - a_{23}x_1^2 \nonumber\\
&+ \frac{b_{12}}{2}x_2 - \frac{b_{13}}{2}x_3 - \frac{b_{22}}{2}x_1 + c_{23}, \\
g_{31} =& - a_{22}x_3x_1 + a_{12}x_2x_3 + a_{23}x_1x_2 - a_{31}x_2^2 \nonumber\\
&+ \frac{b_{23}}{2}x_3 - \frac{b_{21}}{2}x_1 + \frac{b_{11}}{2}x_2 + c_{31} ,
\end{align}
where $a_{ij}, b_{ij}, c_{ij}$ are 20 arbitrary constant parameters.
Then intertwiner Eq.\eqref{16} is rewritten as
\begin{align}
A = a_{ij}l_il_j + b_{ij}l_i\partial_j + c_{ij}\partial_i\partial_j + (d_{ij}x_i + v_j)\partial_j + w,
\end{align}
by introducing (i) $l_i\equiv \epsilon_{ijk}x_j\partial_k$, (ii) $a_{21}, a_{13}, a_{32}$ so that $a_{ij}$ satisfy $a_{ji}=a_{ij}$, (iii) $b_{33}$ so that $\sum_{i=1}^3b_{ii}=0$, (iv) $c_{21}, c_{13}, c_{32}$ so that $c_{ij}$ satisfy $c_{ji}=c_{ij}$ and (v) $d_{ij} \equiv (\sum_{k=1}^3a_{kk})\delta_{ij} - a_{ij}$.

@

We now consider a special solution of Eqs.\eqref{17}, \eqref{18} and \eqref{19}.
We set $a_{ij} = b_{ij} =0$ and $c_{ij} = \mathrm{diag}(1,-1,0)$;
\begin{align}
A = \partial_1^2 - \partial_2^2 + v_i\partial_i + w.
\end{align}
Then Eqs.\eqref{17}, \eqref{18} and \eqref{19} are rewritten as
\begin{align}
&\partial_iv_j + \partial_jv_i = (\widetilde{V}-V)c_{ij},\label{20} \\
&c_{ij}\partial_j(\widetilde{V}+V) + 2\partial_iw = (\widetilde{V}-V)v_i,\label{21} \\
&\boldsymbol{v}\cdot\nabla (\widetilde{V}+V) = 2(\widetilde{V}-V)w.\label{22}
\end{align}
By denoting $x_\pm \equiv (x_1\pm x_2)/\sqrt{2}$, we can write the general solution of Eq.\eqref{20} as follows;
\begin{align}
v_1 &= x_3y_1 + \frac{\beta_+(x_+) + \beta_-(x_-)}{\sqrt{2}}, \\
v_2 &= -x_3y_2 - \frac{\beta_+(x_+) - \beta_-(x_-)}{\sqrt{2}}, \\
v_3 &= a - (cx_+x_- + d_+x_- + d_-x_+) \\
&= a - (\frac{c}{2}x_1^2 + d_1x_1) + (\frac{c}{2}x_2^2 + d_2x_2), \\
\widetilde{V}-V &= 2cx_3 + \beta_+^\prime(x_+) + \beta_-^\prime(x_-),
\end{align}
where $a, c, d_\pm \equiv (d_1\pm d_2)/\sqrt{2}$ are arbitrary constants, $\beta_{\pm}(x_{\pm})$ are arbitrary functions and
\begin{align}
y_{1,2} = \frac{cx_{1,2} + d_{1,2}}{\sqrt{2}}.
\end{align}
By denoting $y_\pm \equiv (y_1 \pm y_2)/\sqrt{2}$, we can write the general solution of Eq.\eqref{21} as
\begin{align}
\widetilde{V}+V &= 2V_3(x_3) + \frac{3}{2}x_3^2(y_+^2+y_-^2) + x_3(y_+\beta_+ + y_-\beta_- + \alpha_2(x_2) - \alpha_1(x_1)) \nonumber\\
&\quad + \frac{\beta_+^2 + \beta_-^2}{2}  + \gamma_2(x_2) - \gamma_1(x_1), \\
2w &= cv_3x_3^2 + v_3(\beta_+^\prime + \beta_-^\prime)x_3 + \beta_+\beta_- + \gamma_1(x_1) + \gamma_2(x_2).
\end{align}
In the above $\alpha_1(x_1)$ and $\alpha_2(x_2)$ must satisfy the condition
\begin{align}
\alpha_1(x_1) + \alpha_2(x_2) = v_3(\beta_+^\prime + \beta_-^\prime) - y_+\beta_- - y_-\beta_+.
\end{align}
Now intertwiner becomes
\begin{align}
A =& 2\left( \partial_+ + \frac{x_3y_+ + \beta_+}{2} \right)\left( \partial_- + \frac{x_3y_- + \beta_-}{2} \right) + v_3\partial_3 + \frac{2cv_3-ca-d_+d_-}{2}x_3^2 \nonumber\\
&+ \frac{v_3(\beta_+^\prime + \beta_-^\prime) - y_+\beta_- - y_-\beta_+}{2}x_3 + \frac{\gamma_1+\gamma_2}{2}.
\end{align}
Here when $v_3=0$ ($\therefore a=c=d_\pm=y_\pm=0$), it reduces to a trivial generalization of two-dimensional case.
Thus we consider the case $v_3\neq 0$.
Then Eq.\eqref{22} is rewritten as four equations;
\begin{align}
&2c\{ 4cv_3 - 3(ca+d_+d_-) \} = 2m_4v_3, \\
&(4c + y_+\partial_+ + y_-\partial_-)\alpha - (ca+d_+d_-)(\beta_+^\prime + \beta_-^\prime) = 2m_3v_3, \\
&\partial_+(\beta_+\alpha + y_+\gamma) + \partial_-(\beta_-\alpha + y_-\gamma) = v_3\{ 2m_2 + 3(y_+^2+y_-^2) \}, \\
&\partial_+(\beta_+\gamma) + \partial_-(\beta_-\gamma) = v_3(2m_1 + y_+\beta_+ + y_-\beta_- + \alpha_2 - \alpha_1)
\end{align}
and
\begin{align}
V_3 = \frac{m_4}{4}x_3^4 + \frac{m_3}{3}x_3^3 + \frac{m_2}{2}x_3^2 + m_1x_3,
\end{align}
where $m$ are constants and $\alpha \equiv \alpha_1(x_1)+\alpha_2(x_2)$, $\gamma \equiv \gamma_1(x_1) + \gamma_2(x_2)$.
We found the general solution of equations for $c\neq0$ as following;
\begin{align}
\beta_\pm &= h_\pm y_\pm + \frac{q_\pm}{y_\pm}, \label{beta} \\
\alpha_1 &= - (\sqrt{2}h_1y_1^2 + \alpha_0), \\
\alpha_2 &= \sqrt{2}h_1y_2^2 + \alpha_0, \\
\gamma_1 &= - \frac{1}{4c^2}\left\{ y_1^4 + (m_2-4c^2h_1^2)y_1^2 + \gamma_0 + \frac{s_1}{y_1^2} \right\}, \\
\gamma_2 &= \frac{1}{4c^2}\left\{ y_2^4 + (m_2-4c^2h_1^2)y_2^2 + \gamma_0 + \frac{s_2}{y_2^2} \right\} ,
\end{align}
where $s, h_\pm \equiv (h_1\pm h_2)/\sqrt{2}, q_\pm \equiv (q_1\pm q_2)/\sqrt{2}$ are arbitrary constants and
\begin{align}
m_4 = 4c^2,\quad m_3= 6\sqrt{2}c^2h_1,\quad m_1 = \frac{h_1(m_2-4c^2h_1^2)}{\sqrt{2}}-\alpha_0,
\end{align}
with conditions
\begin{align}
s_\pm h_2 &= s_\pm q_2 = s_- q_1 = 0. \label{conditions}
\end{align}
Transforming to the standard form as to $x_3$ and omitting constants, we obtain the isospectral potentials;
\begin{align}
V 
&= c^2x_3^4 + \left\{ \frac{3}{4}(y_+^2+y_-^2) + \frac{m_2}{2} - 3c^2h_1^2 \right\}x_3^2
+ \left\{ \frac{h_2}{2\sqrt{2}}(y_+^2-y_-^2) + \frac{q_1}{\sqrt{2}} - c \right\}x_3 \nonumber\\
&+ \frac{1}{8c^2}(y_1^4 + y_2^4) + \left(\frac{m_2}{8c^2}-\frac{3}{4}h_1^2 + \frac{1}{8}h_2^2\right)(y_1^2+y_2^2) + \frac{1}{8c^2}\left(\frac{s_1}{y_1^2}+\frac{s_2}{y_2^2}\right) \nonumber\\
&+\frac{q_+(q_++2c)}{4}\frac{1}{y_+^2} + \frac{q_-(q_-+2c)}{4}\frac{1}{y_-^2} + \frac{1}{\sqrt{2}}ch_1, \\
\widetilde{V}
&= c^2x_3^4 + \left\{ \frac{3}{4}(y_+^2+y_-^2) + \frac{m_2}{2} - 3c^2h_1^2 \right\}x_3^2
+ \left\{ \frac{h_2}{2\sqrt{2}}(y_+^2-y_-^2) + \frac{q_1}{\sqrt{2}} + c \right\}x_3 \nonumber\\
&+ \frac{1}{8c^2}(y_1^4 + y_2^4) + \left(\frac{m_2}{8c^2}-\frac{3}{4}h_1^2 + \frac{1}{8}h_2^2\right)(y_1^2+y_2^2) + \frac{1}{8c^2}\left(\frac{s_1}{y_1^2}+\frac{s_2}{y_2^2}\right) \nonumber\\
&+\frac{q_+(q_+-2c)}{4}\frac{1}{y_+^2} + \frac{q_-(q_--2c)}{4}\frac{1}{y_-^2} + \frac{1}{\sqrt{2}}ch_1.
\end{align}
To repeat, $c,m,h,q,s$ are constants which satisfy the conditions \eqref{conditions} and $y\equiv cx+d$.
Thus we have constructed a class of new three-dimensional isospectral systems, which are forth order in $x_3$.
As these potentials have singularities of inverse square on $y_\pm=0$, we have to make restrictions on the region of definition as $y_\pm > 0$.
These systems have symmetry operator $A^\dag A$ and $AA^\dag$ respectively by virtue of Eq.\eqref{re1} and its Hermitian conjugate.
Unlike one-dimensional case, $A^\dag A$ and $AA^\dag$ are not written in terms of $H$ and $\widetilde{H}$ in general.
Thus the relationship between zero-modes of $A$ and solutions of Schr\"odinger equation with Hamiltonian $H$ is not clear.
This makes the analysis of possible spectral difference more difficult.
We leave it as a future problem.

\section{Conclusion and Outlook}

Using the technique of supersymmetric quantum mechanics, we have constructed a class of three-dimensional isospectral systems.
We have proved that the first order intertwiner case reduce to trivial generalizations of one-dimensional case.
We have constructed new nontrivial three-dimensional systems using the second order intertwiner with special choice of ``metric'' $g_{ij} = \mathrm{diag}(1,-1,0)$.
The constructed isospectral potentials do not contain arbitrary function of $x_3$ but they are forth order in $x_3$
(Of course as a trivial generalization of two-dimensional case, there are solutions which contain arbitrary function of $x_3$).
This limitation of functional form to the forth order happens also in the two-dimensional case with the metric $g_{ij} = \mathrm{diag}(1,0)$\cite{00d}.
We find it interesting that functions $\beta_\pm$ are determined as Eq.\eqref{beta} because of nontrivial coupling to $x_3$ component, though in the two-dimensional case with metric $g_{ij}=(1,-1)$ there are wider classes of functions $\beta_\pm$.
These constructed potentials may not have direct physical siginificance for the present, but they might have some time.
Whether these potentials may find direct physical application or not is left as a future problem.
Even if there is no direct physical application, further analysis of these models will be meaningful because they can be used as a toy model for supersymmetric quantum field theory.
For the same reason, the search for other isospectral systems based on different choices of metric $g_{ij}$ will be interesting.
The problem of possible difference in spectra was left unsolved.
In case of one-dimensional first order SUSY, the condition for missing of ground state is simple.
Yet in case of one-dimensional second order SUSY, the simple condition for spectral difference has not been found.
Whether there is a simple condition or not is not sure.
We might have to deal with it case by case depending on the models.
In case of three-dimensional systems, the situation seems the same.
It is an important problem to be studied for future works.
In this paper, we have made a first step for application of supersymmetric quantum mechanics to construct three-dimensional isospectral systems having only one particle.
The study of three-dimensional system is important because our physical space is three-dimensional, but in general, there are many particles concerning physical phenomenon.
Whether the generalization of this method to the systems of more than two particles is possible or not is an open question.

\section*{Acknowledgements}
I would like to thank Prof. H. Aoyama and Assis. Prof. K. Sugiyama for useful discussions.
I would also like to appreciate Prof. T. Uematsu and Prof. M. Sakagami for support and encouragement.

%

\end{document}